\documentstyle[aaspp4,epsf]{aastex} 

\def\simlt{\mathrel{\spose{\lower 3pt\hbox{$\mathchar''218$}}
     \raise 2.0pt\hbox{$\mathchar''13C$}}}
\def\simgt{\mathrel{\spose{\lower 3pt\hbox{$\mathchar''218$}}
     \raise 2.0pt\hbox{$\mathchar''13E$}}}

\shorttitle{Cloud Disruption in Brown Dwarfs}
\shortauthors{Burgasser et al.}

\begin{document}

\def\gtorder{\mathrel{\raise.3ex\hbox{$>$}\mkern-14mu
             \lower0.6ex\hbox{$\sim$}}}
\def\ltorder{\mathrel{\raise.3ex\hbox{$<$}\mkern-14mu
             \lower0.6ex\hbox{$\sim$}}}

\def\today{\number\year\space \ifcase\month\or  January\or February\or
        March\or April\or May\or June\or July\or August\or
        September\or
        October\or November\or December\fi\space \number\day}
\def\fraction#1/#2{\leavevmode\kern.1em
 \raise.5ex\hbox{\the\scriptfont0 #1}\kern-.1em
 /\kern-.15em\lower.25ex\hbox{\the\scriptfont0 #2}}
\def\spose#1{\hbox to 0pt{#1\hss}}
\def\heion{\ion{He}{2}}
\def\wig#1{\mathrel{\hbox{\hbox to 0pt{%
          \lower.5ex\hbox{$\sim$}\hss}\raise.4ex\hbox{$#1$}}}}
\def\Teff{T_{\rm eff}}
\def\sss{\scriptscriptstyle}

\title{Evidence for Cloud Disruption in the L/T Dwarf Transition}

\author{Adam J.\ Burgasser\altaffilmark{1,2},
Mark S.\ Marley\altaffilmark{3},
Andrew S.\ Ackerman\altaffilmark{3},
Didier Saumon\altaffilmark{4},
Katharina Lodders\altaffilmark{5},
Conard C.\ Dahn\altaffilmark{6},
Hugh C.\ Harris\altaffilmark{6},
and J.\ Davy Kirkpatrick\altaffilmark{7}}

\altaffiltext{1}{UCLA Division of Astronomy and Astrophysics,
8965 Mathematical Sciences Building,
Los Angeles, CA 90095-1562; adam@astro.ucla.edu}
\altaffiltext{2}{Hubble Postdoctoral Fellow}
\altaffiltext{3}{NASA Ames Research Center, MS 245-5, Moffett Field CA, 94035;
mmarley@mail.arc.nasa.gov, ack@sky.arc.nasa.gov} 
\altaffiltext{4}{Department of Physics and Astronomy, Vanderbilt University, 
Nashville, TN 37235; dsaumon@cactus.phy.vanderbilt.edu}
\altaffiltext{5}{Planetary Chemistry Laboratory, Department of Earth
and Planetary Sciences, Washington University, St.\ Louis, MO 63130-4899;
lodders@levee.wustl.edu}
\altaffiltext{6}{United States Naval Observatory, PO Box 1149, Flagstaff,
AZ 86002-1149; dahn@nofs.navy.mil, hch@nofs.navy.mil}
\altaffiltext{7}{Infrared Processing and Analysis Center, MS 100-22, 
770 South Wilson Ave., California Institute
of Technology, Pasadena, CA 91125; davy@ipac.caltech.edu}

\pagestyle{plain}

\begin{abstract}

Clouds of metal-bearing condensates play a critical role in shaping
the emergent spectral energy distributions of the coolest classes of
low-mass stars and brown dwarfs, L and T dwarfs.  
Because condensate clouds in planetary atmospheres show distinct
horizontal structure, we have explored a model for partly cloudy
atmospheres in brown dwarfs.  
Our model successfully
reproduces the colors and magnitudes of both L and T dwarfs for
the first time, including 
the unexpected brightning of the early- and mid-type T dwarfs at J-band,
provided that clouds are rapidly
removed from the photosphere at T$_{eff}$ $\approx$
1200 K.  The clearing of cloud layers also 
explains the surprising persistence and strengthening of gaseous FeH 
bands in early- and mid-type T dwarfs.  The breakup of cloud layers 
is likely driven by convection in the troposphere, analogous to 
phenomena observed on Jupiter.  Our results demonstrate that 
planetary-like atmospheric dynamics must be considered when examining 
the evolution of free-floating brown dwarfs.
\end{abstract}


\section{Introduction}
The recent discovery of a vast population of free-floating brown 
dwarfs in the vicinity of the Sun \citep{kir99,kir00,mrt99,me02,geb02}
has led to the definition 
of two new spectroscopic classes, L dwarfs and T dwarfs.  These are 
the first additions to the standard stellar sequence in over 60 
years, encompassing objects cooler than the M spectral class.  L 
dwarfs \citep{kir99,mrt99} are characterized by absorption bands of metal hydrides 
(e.g., FeH, CrH, and MgH), strong alkali lines (Li, Na, K, Rb, and 
Cs), and H$_2$O and CO absorption in the near-infrared.  Below effective 
temperatures $T_{\rm eff} \sim 1500\,\rm K$, CO reduces to 
$\rm CH_4$ \citep{feg96}, a molecule which has 
distinct absorption bands at 1.6 and $2.2\,\rm\mu m$.  Objects exhibiting 
these $\rm CH_4$ features have grossly different spectroscopic and 
photometric properties as compared to L dwarfs,
and are designated T dwarfs \citep{kir99,kir00,me02}. 
The spectroscopic sequence M 
to L to T is the evolutionary cooling sequence for a typical brown 
dwarf.

L dwarfs are further characterized by the presence of condensates 
near or just below their photospheres.  These condensates include 
liquid iron; solid VO; and aluminum, calcium, 
magnesium, and titanium-bearing minerals such as enstatite ($\rm MgSiO_3$), 
grossite ($\rm CaAl_4O_7$), and perovskite ($\rm CaTiO_3$) \citep{bur99,lod99}.   
The opacity of 
the condensate grains and droplets gives L dwarfs fairly red 
near-infrared colors, with $ J-K_s \sim 2$ \citep{kir00}.
Conversely, condensates in mid- 
and late-type T dwarfs appear to lie deep below the photosphere. 
These objects have neutral near-infrared colors, $ J-K_s \sim 0$, governed 
by strong absorption features of $\rm H_2O$, $\rm CH_4$, and 
collision-induced $\rm H_2$, 
particularly at K-band \citep{me02}.  

Three classes of models have been developed 
to account for condensates in brown dwarf atmospheres.  The dusty 
model \citep{lun89,cha00,all01} assumes that condensates remain dispersed in the gas and in 
chemical equilibrium throughout the atmosphere.  The clear model \citep{bur97} 
removes all condensates from the photosphere as they form, presumably 
through gravitational settling.  The cloudy model \citep{ack01,mar02} 
incorporates 
condensate cloud formation based on an assumed sedimentation 
efficiency, $f_{\rm rain}$.  The dusty and clear models are extreme cases of 
the cloudy model with very low and very high values of $f_{\rm rain}$, 
respectively.  

In this paper, we
present an exploratory model of partial cloud clearing in cool dwarf
atmospheres derived from the cloudy and clear models. 
In $\S$2 we describe 
our model and compare its derived photometry to 
the near-infrared color-magnitude diagram of late-type
dwarfs.  This analysis indicates that a rapid clearing of clouds
takes place across the L/T transition.
In $\S$3 we present further evidence of cloud clearing in the
persistance and strengthening 
of FeH absorption at 9896 {\AA} in the early- and mid-type T dwarfs.  We
discuss our results in $\S$4.

\section{Photometric Evidence for Cloud Clearing}

\subsection{The Color-Magnitude Diagram for Late-type Dwarfs}

\placefigure{fig-1}

Figure 1 shows the near-infrared color-magnitude 
diagram for a sample 
of M, L, and T dwarfs with known parallaxes.  Photometry were obtained
from the Two Micron All-Sky Survey \citep[hereafter 2MASS]{skr97}
and parallaxes from the USNO Parallax Program \citep{dah02} or from
Hipparcos \citep{pry97} in the case of companions to nearby stars.
As described in \citet{dah02}, late-type M and L dwarfs are increasingly redder
and fainter toward later spectral types, but the transition to the
bluer T dwarfs is marked by a distinct brightening at J-band, with both 
the T2 dwarf SDSS J1254-0122 \citep{leg00} and the T5 dwarf 2MASS 
0559-1404\footnote{\citet{me01} has suggested that 2MASS J0559-1404 may be
an unresolved binary brown dwarf. 
However, even if it were an equal-magnitude double,
it would still remain brighter at J-band than the latest L
dwarfs.} \citep{me00} being brighter at J-band 
than the latest-type L dwarfs.  Note that neither of these objects
is expected to be young and hence less compact (R $\gtrsim$ 1 R$_{Jup}$;
Burrows et al.\ 1997) based on extremely weak or absent H$\alpha$ 
emission (Burgasser et al. 2002a; Kirkpatrick et al., in preparation).
Late-type T dwarfs
are increasingly bluer in $J-K_s$ color and fainter at J-band.  

Also shown in this figure are 
color-magnitude tracks for the three atmosphere models described above. 
The dusty model \citep[dashed line]{cha00} reproduces the colors of 
early-type L dwarfs, but continues 
to redden to colors far greater than those observed for the mid- and 
late-type L dwarfs.  This is due to the continued influence of
dust opacity, which mutes the otherwise strong absorption bands of
H$_2$O, CH$_4$, and H$_2$.  The clear model (solid line, left)
reproduces the colors of mid- 
and late-type T dwarfs, but not those of L dwarfs.  Finally, the cloudy model 
(solid line, right)
successfully reproduces the near-infrared color limit for the 
latest-type L dwarfs, $J-K_s \lesssim 2$-$2.5$ \citep{kir00,mar02}, and subsequently
turns to bluer near-infrared colors.  However, the predicted
transition occurs over a greater 
decrease in brightness, $\Delta M_J = 2-3$ mag from 
$J-K_s = 2$ to $J-K_s = 0$, than is observed.
Increasing $f_{\rm rain}$ in this model (i.e., moving toward a more cleared
photosphere) allows the color 
transition to occur at brighter magnitudes, but fails to match the 
colors and magnitudes of the latest L dwarfs.  Therefore, the 
condensate cloud model, while adequately reproducing the brightness and color 
evolution of L dwarfs, cannot reproduce the observed rapid 
transition to the T dwarfs.

Is it possible to resolve this apparent discrepancy between the 
cloudy model and the observations?  All three atmosphere models assume 
horizontal homogeneity and therefore uniform condensate distribution 
over the visible disk of the brown dwarf.  However, clouds in the 
planets of the solar system, for instance Jupiter's upper cloud decks 
of solid NH$_3$ and NH$_4$SH, display discrete 
structures of spots and bands caused by tropospheric weather 
patterns.  Breaks between the clouds on Jupiter coincide with bright 
``hot-spots'' at 5 $\micron$ \citep{wes74},
as light emerges from warmer regions well 
below the upper cloud decks.  Since horizontally non-homogeneous clouds 
are ubiquitous in planetary atmospheres, it is reasonable to expect 
that similar features could be present in the condensate cloud layers of 
brown dwarfs as well \citep{ack01}.

\subsection{Cloud Clearing Model}

Connecting the cloudy and clear tracks in Figure 1
are the predicted fluxes for an alternative model we have 
developed that allows for partial clearing of condensate clouds.
These were constructed by 
first linearly interpolating between the fluxes of the clear model 
and a modified cloudy model in which the cloud opacity has been 
removed, weighing by the fraction of cloud coverage.  This 
combination approximates the effect of partial cloud opacity on the 
underlying atmospheric pressure-temperature profile.  We then 
linearly interpolated the resulting fluxes with those of the standard 
cloudy model, again weighing by the fractional cloud coverage, to 
derive the relative brightness contributions of cloudy and cloud-free 
regions. Our simple approach assumes that many clouds are distributed 
randomly across the observed disk.  

When compared to the data, we see that the evolution from the red and 
dusty L dwarfs to the blue and clear T dwarfs can be traced (heavy 
line) down the cloudy track to an effective temperature 
$T_{\rm eff} \approx 1200\,\rm K$, across the 
1200 K partly cloudy model, and subsequently down the cloud-free track. 
SDSS J1254-0122 is a true transition object, lying close to the 1200 K partly 
cloudy track at roughly 40\% cloud coverage.  2MASS J0559-1404, 
however, appears to have a cloud-free photosphere.  
The T$_{eff}$ at which the color transition appears to 
take place is only slightly cooler than that estimated 
for one of the latest-type L dwarfs, Gliese 584C
($\Teff = 1350\,\rm K$; Kirkpatrick et al.\ 2000; Dahn et al.\ 2002).
The clearing of clouds at 
nearly constant T$_{eff}$ explains the apparent 
brightening of the early-type T dwarfs at 1 $\micron$, as the emitting 
layers below the clouds in these objects would be warmer than the 
layers above the cloud decks in late-type L dwarfs.
Note that there is 
some intrinsic scatter among the empirical data, likely 
due to the influence of gravity, rotation, and metallicity on cloud 
coverage\footnote{Magnetic activity may also be a source of 
photometric scatter.  Magnetically active low-mass dwarfs have
extended cool spots \citep{ale97} which could exhibit
stronger H$_2$O and/or CH$_4$ absorption features, resulting in later-type
spectral morphologies and photometric colors for a given T$_{eff}$.  
This phenomenon would not explain the rapid L/T transition, however, 
again due to the lack of strong
H$\alpha$ emission in either SDSS J1254-0122 or 2MASS J0559-1404, 
as well as the very low photospheric magnetic Reynolds numbers
in these cool, neutral atmospheres \citep{gel02}.}.
Nonetheless, the 
apparently sudden color transition from L to T clearly cannot
be explained by the progressive sinking of a uniform cloud deck with 
decreasing $T_{\rm eff}$, but is possible 
through the disruption of condensate clouds. 

\section{Spectroscopic Evidence for Cloud Clearing: FeH Absorption}

There is additional evidence that light emitted at $1\,\rm \mu m$ by early and 
mid-type T dwarfs is seen through opening holes in condensate cloud decks. 
Figure 2 plots the optical spectra of a series of late-type L and T 
dwarfs (Kirkpatrick et al.\ 1999, 2000; Burgasser et al., in preparation),
obtained using the Low Resolution Imaging 
Spectrograph \citep{oke95} on Keck.  Spectral types listed are
from \citet{kir99,kir00} and \citet{me02}, and are consistent with a 
monotonic T$_{eff}$ sequence \citep{me02}.   Here we 
focus on the region centered at the $9896\,\rm \AA$ FeH band.  This 
feature weakens in the later-type L dwarfs, in concert with the 
observed disappearance of the $8962\,\rm \AA$ FeH band \citep{kir99}.
This is 
consistent with the segregation of liquid iron into condensate 
clouds, which depletes the atmosphere of iron-bearing gases above the 
cloud layer.  As the clouds settle below the photosphere with 
decreasing temperature, the amount of detectable FeH drops as well. 
However, the $9896\,\rm \AA$  FeH band unexpectedly reappears in SDSS J1254-0122 
and is quite pronounced in 2MASS J0559-1404.  The retention of FeH 
absorption from late-type L dwarfs to early- and mid-type T dwarfs is 
a general trend, as shown in Figure 3.  The FeH band clearly weakens 
from L4 to L8, but persists and even strengthens slightly toward type 
T5.5, subsequently weakening in the latest T dwarfs.  

\placefigure{fig-2}
\placefigure{fig-3}

The plateau in 
band strength across the L/T transition cannot be explained with a 
cloud deck that gradually sinks below the photosphere with decreasing 
T$_{eff}$.  In this case, we would only see the 
atmosphere above the cloud deck in which FeH is depleted.  However, 
if holes are present in the clouds, deeper and warmer regions in 
which FeH is not depleted become observable.  These deep layers can 
be detected at $1\,\rm \mu m$, where $\rm H_2O$, $\rm CH_4$, and $\rm H_2$ 
absorption is relatively 
weak.  Since the reappearance of FeH coincides with the turnover in 
$J-K_s$ color, the hypothesis of cloud clearing simultaneously 
explains both phenomena.

\section{Discussion}

The disruption of condensate clouds in brown dwarf atmospheres likely 
begins when the cloud deck forms within the atmospheric 
convection zone and becomes subject to coherent vertical motions.  At 
higher effective temperatures, the clouds lie primarily above the 
radiative-convective boundary and would likely remain spatially 
uniform, perhaps like the stratospheric hazes on Titan and the giant 
planets.  Evidence for vertical transport has already been seen in cool brown 
dwarfs, with the detection of CO absorption at $4.7\,\rm\mu m$ in the T6.5 
dwarf Gliese 229B \citep{nak95} in excess of its equilibrium abundance
\citep{nol97,opp98}. 
Photometric studies of late-type L and T dwarfs also indicate far 
greater absorption in the $5\,\rm \mu m$ band than predicted by chemical 
equilibrium models \citep{leg02,rei02}, likely caused by enhanced CO abundances 
brought up by vertical flows.  Since the principal cloud decks lie 
within 0.5 pressure scale heights of one another, the convective 
motion need not be particularly vigorous to disrupt all layers.  Note 
that upwelling brings substantial CO into the photosphere because of 
its stable bond and hence relatively slow conversion to $\rm CH_4$ 
\citep{lod02}.  The 
fragility of the FeH bond (1.63 eV versus 11.09 eV for CO) precludes 
any similar non-equilibrium enhancement of this species above the 
cloud layer.

The picture of a brown dwarf's atmospheric evolution presented here 
has a number of observable consequences.  First, the 
T$_{eff}$ scale between the late-type L and early-type T dwarfs 
should be very narrow; this has already been surmised by a number of 
studies \citep{kir00,me02}.
A narrow temperature scale also implies fewer early-type 
T dwarfs than mid- or late-type T dwarfs, given the shorter time a 
brown dwarf spends in this phase of its thermal evolution.  The 
statistics are currently insufficient to adequately test this 
prediction.  We also expect substantial photometric and/or 
spectroscopic variability in late-type L and early-type T dwarfs, 
caused by the formation, evolution, and motion (rotational or 
wind-driven) of cloud holes in their upper atmospheres.  Such weather 
phenomena have been previously cited to explain non-periodic 
variability in late-type M and early-type L dwarfs \citep{bai01,mrt01}. 
Similar 
variability could be significant at $1\,\rm \mu m$ in the L/T transition. 

Filling in the color-magnitude sequence of late-type L and early-type 
T dwarfs will enable us to further characterize the behavior of 
condensate clouds in brown dwarf atmospheres.  Regardless, the 
observed spectral and photometric properties of brown dwarfs indicate 
that the planetary-like phenomena of cloud formation and disruption 
are important in the evolution of free-floating brown dwarfs.

\acknowledgements

We thank A. Burrows and I. McLean for valuable discussions, M. 
Putman and K. Zahnle for assistance in the preparation of the 
manuscript, and G.\ Chabrier for an electronic version of his dusty models
in the 2MASS J and K$_s$ bands.  We also thank our anonymous referee for
bringing up the issue of magnetic activity in the L/T transition.
AB acknowledges support by NASA through Hubble Fellowship grant
HST-HF-01137.01 awarded by the Space Telescope Science Institute,
which is operated by the Association of Universities for Research in
Astronomy, Inc., for NASA, under contract NAS 5-26555.  MM
acknowledges support from NSF grant AST 00-86288 and NASA grant
NAG5-9273.  Work by KL and DS is supported by NSF grant AST 00-86487
and NASA grant NAG5-4988 respectively.
Portions of the data presented herein were obtained at 
the W. M. Keck Observatory which is operated as a scientific 
partnership among the California Institute of Technology, the 
University of California, and the National Aeronautics and Space 
Administration.  The Observatory was made possible by the generous 
financial support of the W. M. Keck Foundation.  This publication 
makes use of data from the Two Micron All Sky Survey, which is a 
joint project of the Unversity of Massachusetts and the Infrared 
Processing and Analysis Center, funded by the National Aeronautics 
and Space Administration and the National Science Foundation.

\clearpage

\figcaption[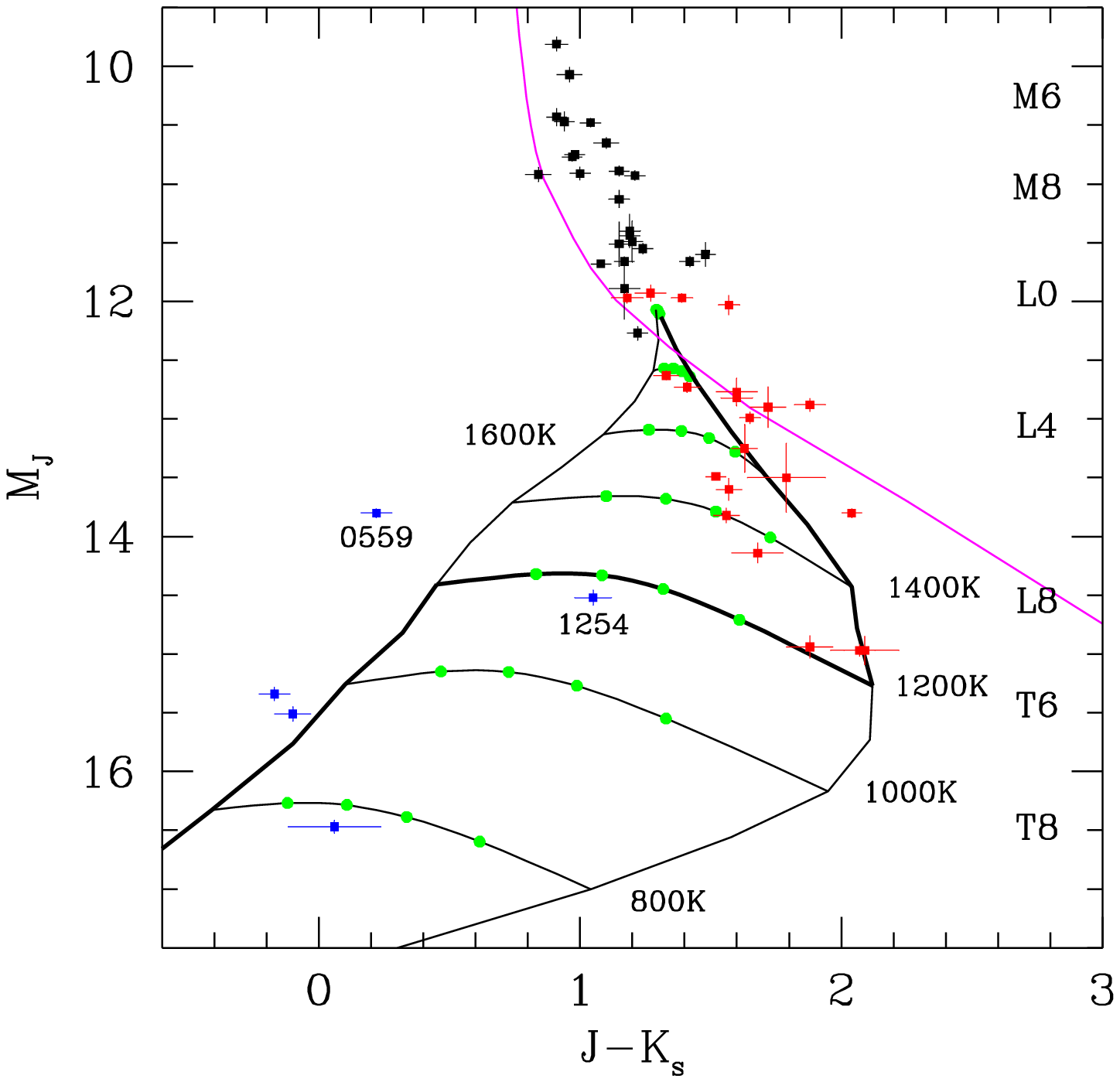]{Near-infrared color-magnitude diagram of M, L, and T
dwarfs.  Absolute J magnitudes and $J-K_s$ infrared colors are shown for 
a sample of M (black), L (red), and T (blue) dwarfs with known 
parallaxes.  The positions of 2MASS J0559-1404 and SDSS J1254-0122 are
indicated.  The predicted colors and magnitudes for the dusty (magenta),
clear (black, left), and cloudy (black, right; $f_{rain}$ = 3)
atmosphere models are plotted as a 
function of T$_{eff}$ at constant gravity, $g$ = 10$^5$ cm s$^{-2}$,
(typical for very 
low-mass main sequence stars and evolved brown dwarfs).  
Connecting the cloudy 
and clear tracks are the predicted fluxes for our partly 
cloudy models at T$_{eff}$ = 800, 1000, 1200, 1400, 1600, and 1800 K. 
Green dots indicate cloud 
coverage fraction in steps of 20\%.
The apparent evolutionary track of brown dwarfs based 
on the empirical data is indicated by the thickened line, which 
crosses from the cloudy to clear track at T$_{eff}$ $\approx$ 1200 K.  
\label{fig-1}}

\figcaption[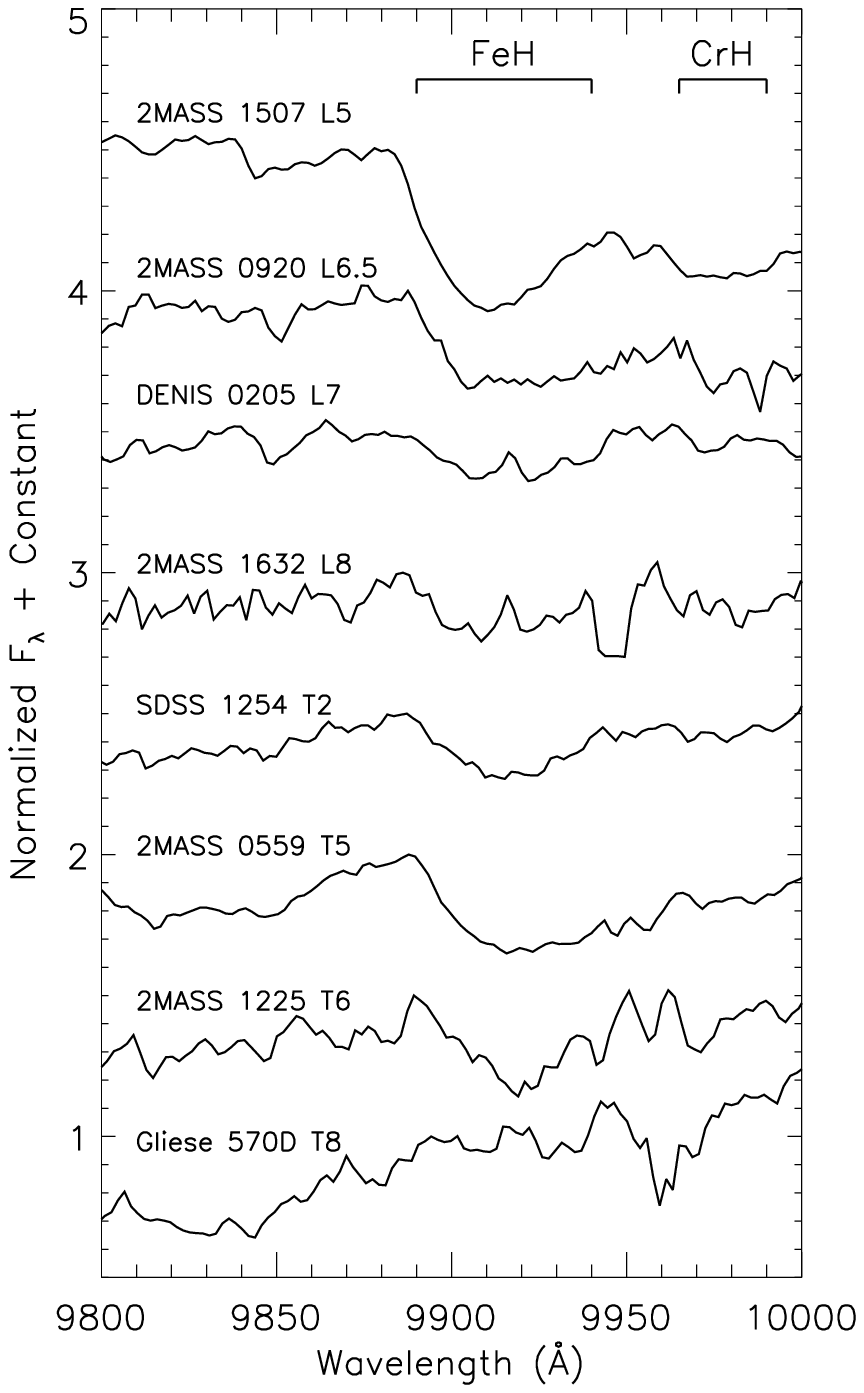]{Evolution of FeH with spectral type.  Optical spectra of 
four late L dwarfs (L5 2MASS J1507-1627, L6.5 2MASS J0920+3517, L7 
DENIS J0205-1159, and L8 2MASS J1632+1904) and four T dwarfs (T2 SDSS 
J1254-0122, T5 2MASS J0559-1404, T6 2MASS J1225-2739, and T8 Gliese 
570D) are shown in the 9800-10000$\,\rm \AA$ regime.  Spectral types
are from \citet{kir99,kir00} and \citet{me02}.
Spectra are normalized 
at 9870 {\AA} and offset for clarity.  The 9896 {\AA} FeH and 9969 {\AA} CrH
bands are indicated.  
The FeH band rapidly fades from 
L5 to L8, but appears to recover and 
strengthen toward T5, disappearing again in the coolest T dwarfs.
\label{fig-2}}

\figcaption[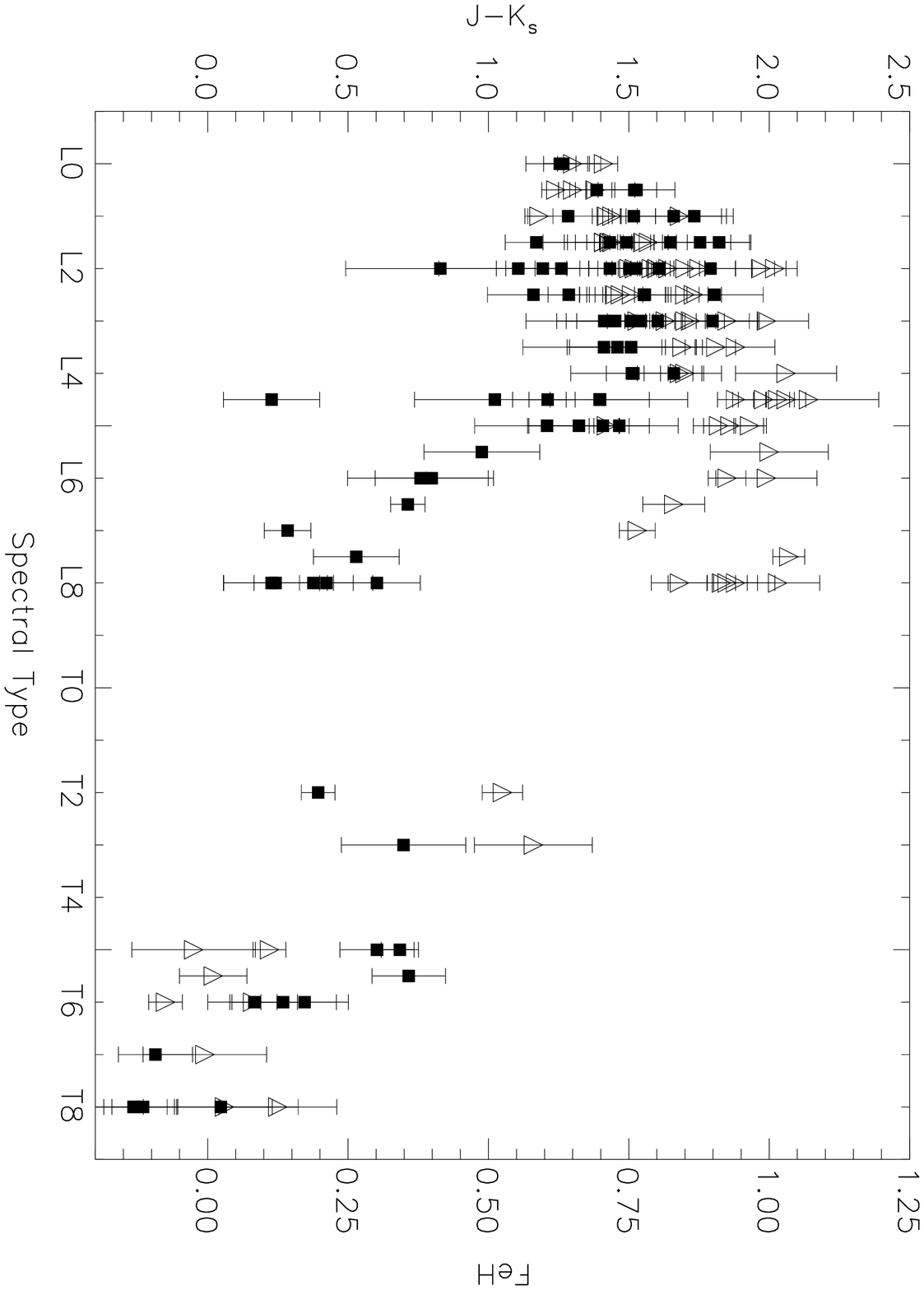]{FeH band strengths and near-infrared colors as a function 
of spectral type.  Solid squares plot the $9896\,\rm \AA$ FeH band index
\citep{kir99} of L 
and T dwarfs measured from optical spectra, while open triangles plot 
their 2MASS $J-K_s$ colors.  Spectral types are from \citet{kir99,kir00} and
\citet{me02}.  The FeH band weakens rapidly from the mid- 
to late-type L dwarfs, where $J-K_s$ becomes slightly redder before 
leveling off at $J-K_s \sim 2$.  Instead of disappearing, however, the 
FeH band persists through T5.5, with some objects showing slightly 
stronger FeH absorption than the latest L dwarfs.
\label{fig-3}}

\clearpage

\begin{figure}
\epsscale{1.1}
\plotone{f1.eps}
\end{figure}

\clearpage

\begin{figure}
\epsscale{0.8}
\plotone{f2.eps}
\end{figure}

\clearpage

\begin{figure}
\epsscale{0.9}
\plotone{f3.eps}
\end{figure}

\end{document}